# Hurwitz's matrices, Cayley transformation and the Cartan-Weyl basis for the orthogonal groups

**M. Hage Hassan**
Université Libanaise, Faculté des Sciences Section (1)
Hadath-Beyrouth

**Abstract**

We find the transformations from the basis of the hydrogen atom of n-dimensions to the basis of the harmonic oscillator of N=2(n-1) dimensions using the Cayley transformation and the Hurwitz matrices. We prove that the eigenfunctions of the Laplacian $\Delta_n$ are also eigenfunctions of the Laplacien $\Delta_N$ for n=1, 3, 5 and 9. A new parameterization of the transformation $R^8 \to R^5$ is derived.
This research leads us first to a new class of spherical functions of the classical groups we call it the bispherical harmonic functions. Secondly: the development of Hurwitz's matrix in terms of adjoint representation of the Cartan-Weyl basis for the orthogonal groups SO(n) leads to what we call the generating matrices of the Cartan-Weyl basis and then we establish it for $n = 2^m (m = 1,2…)$.

**Résumé**

Nous trouvons les transformations de la base de l'atome d'hydrogène de dimension n à la base de l'oscillateur harmonique de dimension N=2 (n-1) en utilisant la transformation de Cayley et les matrices de Hurwitz . Nous montrons que les fonctions propres du Laplacian $\Delta_n$ sont également fonctions propres du Laplacien $\Delta_N$ pour n=1,3, 5 et 9. Une nouvelle paramétrisation de la transformation est dérivée. Cette recherche nous mène d'abord à une nouvelle classe des fonctions sphériques des groupes classiques que nous l'appelons les fonctions harmoniques bisphériques. Deuxièmement : le développement de la matrice de Hurwitz en termes de la représentation adjointe de la base de Cartan-Weyl pour les groupes orthogonaux SO(n) mène à ce que nous appelons les matrices génératrices de la base de Cartan-Weyl et nous l'établissons pour $n = 2^m (m = 1,2…)$.

## 1. Introduction

The link between the Kepler problem and the oscillation is introduced by Binet in classical mechanic. In Quantum mechanic the link between the N=2(n-1) dimensional harmonic oscillator and the n-dimensional hydrogen atom was known to the physists since Schrödinger. The transformation (K-S) introduced by Kustaanheimo-Steifel [1] in celestial mechanics was used by many authors [2-3] for the connection of $R^3$ hydrogen atom and $R^4$ harmonic oscillator. After that many papers [4-5] were devoted to the generalization of this transformation and the conformal transformation of Levi-Civita [6], using Hurwitz's matrix, which was introduced for the solution of the problem: the sum of squares.

These transformations continue to be interesting for their potential relevance to physics. However, its relation with the orthogonal groups was not emphasized, in spite of the well-known work of Weyl [7] on the transformation of Cayley and the group SO(3). I have not found any extension of this work; it seems to me natural to investigate in this direction. I noticed that the multiplication of the orthogonal matrix $O_n = \frac{I - S_n}{I + S_n}$ by a reel number $r \geq 0$, choosing the anti-symmetric matrix $rS_n$, as the principal minor of Hurwitz matrix, leads to components of the last row or the last column function of N=2(n-1) parameters. But we know [8, 9] that last row or the last column of the matrix of the orthogonal groups are the spherical coordinates on the unitary sphere $S_{n-1}$ therefore the number of parameters is n -1. We deduce consequently that there is a transformation, the Hurwitz transformation, which corresponds well to our criterion. We obtain these transformations by two methods: the first one is a direct calculus using a symbolic computer program the second is an analytic method for the calculation of the left column and right row.

One of the goals of these transformations is to find the harmonic spherical functions eigenfunction of the Laplacian $\Delta_n$ which are also eigenfunction of the Laplacian $\Delta_N$. In particular cases we prove a generalization of these harmonic spherical functions not just for n=1, 3 and 5 but also for n=9. We derive then a new parameterization of the transformation $R^8 \to R^5$ using the representation matrix of the group SO (4). We will prove the existence of the generalizations of the spherical functions left and right [8] for the classical groups .we call these functions the bispherical harmonic functions. Considering the importance of these functions in mathematical physics we will study it in another paper.

The adjoint representations of the orthogonal groups are anti-symmetric and the number of elements is n (n-1)/2. The matrix $S_n$ is anti-symmetric and function of (n-1) parameters, {u}, and develops in terms of the adjoint representation of SO (n) [10]. To generate the Cartan-Weyl basis we need consequently n/2 matrices, this number is in agreement with the number of the simple roots of the orthogonal groups [10].By analogy with the generating functions we call these matrices by the generating matrices of the Cartan-Weyl basis and we build it for the cases $n = 2^m$ $(m = 1, 2, \ldots)$..

The first part is devoted for the derivation of Hurwitz transformations. The bispherical harmonic functions and the parameterization of the transformation $R^8 \to R^5$ are the subject of the second part. The introduction of the generating matrices of Cartan-Weyl basis will be treated in third part. The appendix is reserved to the tables of the generators of SO (5). We emphasize that this work is the continuation of our preceding paper [11].

## 2. Hurwitz's matrix and the Cayley transformation.

For the derivation of the orthogonal matrices we take the miners of Hurwitz matrix and we use a computer symbolic program. Then we expose in analytic method for the calculation of the last row and column of these matrices.

## 2.1 The Hurwitz matrix

Let us consider one of the orthogonal matrices [12] of the 8x8 real matrices $H_8 = (h_{ij})$:

$$H_8 = \begin{pmatrix} u_1 & u_2 & u_3 & u_4 & u_5 & u_6 & u_7 & u_8 \\ -u_2 & u_1 & u_4 & -u_3 & u_6 & -u_5 & -u_8 & u_7 \\ -u_3 & -u_4 & u_1 & u_2 & u_7 & u_8 & -u_5 & -u_6 \\ -u_4 & u_3 & -u_2 & u_1 & u_8 & -u_7 & u_6 & -u_5 \\ -u_5 & -u_6 & -u_7 & -u_8 & u_1 & u_2 & u_3 & u_4 \\ -u_6 & u_5 & -u_8 & u_7 & -u_2 & u_1 & -u_4 & u_3 \\ -u_7 & u_8 & u_5 & -u_6 & -u_3 & u_4 & u_1 & -u_2 \\ -u_8 & -u_7 & u_6 & u_5 & -u_4 & -u_3 & u_2 & u_1 \end{pmatrix} \quad (1)$$

Where $u_i$ (i =1, 1, 8) are real numbers. In order to generate Cayley transformation we assign $H_n$ to the formed matrix of first n line and n column, and $S_n$ is the same matrix with $u_1 = 0$. It can be shown that the matrix $H_8$ can be developed as a linear combination of Clifford matrices. Indeed, we have

$$H_8 = u_1 I - \sum_{i=2}^{8} u_i \Gamma_i^t \quad (2)$$

Where $I$ the identity matrix, t is the transpose and the matrices $\Gamma_i$ satisfy

$$\Gamma_i \Gamma_j + \Gamma_j \Gamma_i = -2I, \ (i, j = 2, \ldots, 8) \quad (3)$$

## 2.2 Cayley Transformation and Hurwitz's matrices

The Cayley transformation for the orthogonal groups $O_n$ is:

$$O_n = \frac{I - S_n}{I + S_n} \quad (4)$$

$S_n$ Is a skew symmetric matrix of order n.
In order to obtain $O_n$ in terms of the variables {u}, we multiply the numerator and denominator by $u_1$ and $O_n$ by $r_n$.

$$r_n = \vec{u}^2 = \sum_{i=1}^{N} u_i^2$$

To simplify the notation we replace $u_1 S_n$ by $S_n$ in the expression of $O_n(u)$ we obtain

$$O_n(u) = \vec{u}^2 \frac{u_1 I - S_n}{u_1 I + S_n} \quad (5)$$

### 2.2.1 Transformation $R^2 \to R^2$

For n=2 we have

$$S_2 = \begin{pmatrix} 0 & u_2 \\ -u_2 & 0 \end{pmatrix}$$

And simple calculation gives

$$O_2 = \frac{1}{r_2}\begin{pmatrix} u_1^2 - u_2^2 & -2u_1u_2 \\ 2u_1u_2 & u_1^2 - u_2^2 \end{pmatrix} \quad (6)$$

### 2.2.2 Transformation $R^4 \to R^3$

Using a computer symbolic program we find the Weyl's expression

$$O_3(u) = (r_3 I - 2u_1 S_3 + 2S_3^2)$$

$$= \begin{pmatrix} u_1^2 - u_2^2 - u_3^2 + u_4^2 & -2(u_2u_1 + u_3u_4) & -2(u_3u_1 - u_2u_4) \\ 2(u_2u_1 - u_4u_3) & u_1^2 - u_2^2 + u_3^2 - u_4^2 & -2(u_1u_4 + u_2u_3) \\ 2(u_3u_1 + u_2u_4) & 2(u_1u_4 - u_2u_3) & u_1^2 + u_2^2 - u_3^2 - u_4^2 \end{pmatrix} \quad (7)$$

In the space of 4-dimensions we derive also the expression

$$\begin{pmatrix} (H_3^t) & V_3^t \\ -V_3 & u_1 \end{pmatrix}\begin{pmatrix} (H_3^t) & -V_3^t \\ V_3 & u_1 \end{pmatrix} = \begin{pmatrix} O_3(u) & 0_3^t \\ 0_3 & \vec{u}^2 \end{pmatrix} \quad (8)$$

With $\quad V_3 = (u_4 \ -u_3 \ u_2)$ and $0_3 = (0 \ 0 \ 0)$.

### 2.2.3 Transformation $R^8 \to R^7$

We also find, using the symbolic program, an analogue expression as above-mentioned:

$$(O_7) = \frac{1}{r_5}(r_5 I - 2u_1 S_7 + 2S_7^2) \quad (9)$$

And in the 8-dimensions space we derive the expression

$$\begin{pmatrix} (H_7^t) & V_7^t \\ -V_7 & u_1 \end{pmatrix}\begin{pmatrix} (H_7^t) & -V_7^t \\ V_7 & u_1 \end{pmatrix} = \begin{pmatrix} O_7(u) & 0_7^t \\ 0_7 & \vec{u}^2 \end{pmatrix} \quad (10)$$

With $\quad V_7 = (u_8 \ u_7 \ -u_6 \ -u_5 \ u_4 \ u_3 \ -u_2)$

And $\quad 0_7 = (0 \ 0 \ 0 \ 0 \ 0 \ 0 \ 0)$.

## 2.3 Calculation of the last column and row of $R^8 \to R^5$ and $R^{16} \to R^9$

We give the method for the calculation of the last column and row using the system of equations

$$(u_1 I + S_n)O_n(u) = \vec{u}^2 (u_1 I - S_n) \quad (11)$$
$$O_n(u)(u_1 I + S_n) = \vec{u}^2 (u_1 I - S_n) \quad (12)$$

Put $O_n(u) = (x_{ij})$ for n= 3, 5, 9, and

$$\rho_2^2 = \sum_{i=n}^{N} u_i^2, \quad \rho_1^2 = \sum_{i=0}^{n-1} u_i^2 \quad (13)$$

We can say left and right with respect to the diagonal of the matrix $O_n$ instead of line and row.

### 2.3.1 Expression of the last column

After doing the identification of the two sides of the first (n-1) equations of the

(11) System and replacing $x_{nn}$ by $\rho_2^2 - \rho_1^2$ in last raw we obtain:

$$H_{(n-1)} \begin{pmatrix} x_{1,n} \\ \vdots \\ x_{(n-1),n} \end{pmatrix} = 2 \begin{pmatrix} h_{n,1} \\ \vdots \\ h_{n,(n-1)} \end{pmatrix}$$

The matrix $H_{(n-1)}$ is orthogonal therefore we find:

$$\begin{pmatrix} x_1^R \\ \vdots \\ x_{n-1}^R \end{pmatrix} = \begin{pmatrix} x_{1,n} \\ \vdots \\ x_{(n-1),n} \end{pmatrix} = 2H_{(n-1)}^t \begin{pmatrix} h_{n,1} \\ \vdots \\ h_{n,(n-1)} \end{pmatrix} \quad (14)$$

### 2.3.2 Expression of the last column

After doing the identification of the two sides of the first (n-1) equations of the (12) System and replacing $x_{nn}$ by $\rho_2^2 - \rho_1^2$ in last raw we obtain:

$$H_{(n-1)}^t \begin{pmatrix} x_{n,1} \\ \vdots \\ x_{n,(n-1)} \end{pmatrix} = 2 \begin{pmatrix} h_{1,n} \\ \vdots \\ h_{(n-1),n} \end{pmatrix}$$

The matrix $H_{(n-1)}$ is orthogonal therefore we deduce:

$$\begin{pmatrix} x_1^L \\ \vdots \\ x_{n-1}^L \end{pmatrix} = \begin{pmatrix} x_{n,1} \\ \vdots \\ x_{n,(n-1)} \end{pmatrix} = 2H_{(n-1)}^t \begin{pmatrix} h_{1,n} \\ \vdots \\ h_{(n-1),n} \end{pmatrix} \quad (15)$$

### 3. The bispherical harmonic functions

We prove that the Laplacien $\Delta_{u,N}$ generate the harmonic functions left and right, and then we treat the parameterization of the transformation $R^8 \to R^5$. We show also the existence of bispherical functions of the classical groups.

### 3.1 The Laplacien of bispherical functions for SO(n), n=2, 3, 5 and 9.

We want to prove that the Laplacien $\Delta_{u,N}$ generate the harmonic functions left and right or simply the bispherical functions not simply for n=3, 5 as well known [4] but also for the case n=9.

We have $\quad \dfrac{\partial f(x^R)}{\partial u_i} = \sum_{l=0}^{n} \dfrac{\partial x_l^R}{\partial u_i} \dfrac{\partial f(x^R)}{\partial x_l^R} + \dfrac{\partial x_{n+1}^R}{\partial u_i} \dfrac{\partial f(x^R)}{\partial x_{n+1}^R}$

With the help of the relations $\quad \dfrac{\partial^2 x_l^R}{\partial u_i^2} = 0, (l \leq n)$ and $\dfrac{\partial x_{n+1}^R}{\partial u_i} = \{ \begin{array}{l} 2u_i \text{ if } i \leq n \\ -2u_i \text{ if } i > n \end{array}$

We can write

$$\frac{\partial^2 f(x^R)}{\partial u_i^2} = \sum_{k,l=0}^{n} \frac{\partial x_k^R}{\partial u_i} \frac{\partial x_l^R}{\partial u_i} \frac{\partial^2 f(x^R)}{\partial(x_l^R)\partial(x_k^R)} + \sum_{l}^{n} \frac{\partial x_l^R}{\partial u_i} \frac{\partial x_{n+1}^R}{\partial u_i} \frac{\partial^2 f(x^R)}{\partial(x_l^R)\partial(x_{n+1}^R)}$$

$$+ \frac{\partial^2 x_{n+1}^R}{\partial u_i^2} \frac{\partial f(x^R)}{\partial x_{n+1}^R} + (\frac{\partial x_{n+1}^R}{\partial u_i})^2 \frac{\partial f(x^R)}{\partial x_{n+1}^R}$$

From $\quad x_n^R = \rho_2^2 - \rho_1^2$

We deduce that $\quad \sum_{i=0}^{N} \frac{\partial^2 x_{n+1}^R}{\partial u_i^2} = 0$, and $\sum_{i=0}^{N} (\frac{\partial x_{n+1}^R}{\partial u_i})^2 = 4\vec{u}^2$

$x_{n1}^R$ Is a homogenous function in terms of $u_i$ we derive that

$$\sum_{i=1}^{N} \frac{\partial x_l^R}{\partial u_i} \frac{\partial x_{n+1}^R}{\partial u_i} = \sum_{i=1}^{n+1} 2u_i \frac{\partial x_l^R}{\partial u_i} - \sum_{i=n+2}^{N} 2u_i \frac{\partial x_l^R}{\partial u_i}$$

$$= 2x_l^R - 2x_l^R = 0$$

The matrix $H_n$ is orthogonal then we deduce

$$\sum_{i=1}^{n} \frac{\partial x_k^R}{\partial u_i} \frac{\partial x_l^R}{\partial u_i} = 4\rho_1^2 \delta_{k,l}$$

$$\sum_{i=n+1}^{2n} \frac{\partial x_k^R}{\partial u_i} \frac{\partial x_l^R}{\partial u_i} = 4\rho_2^2 \delta_{k,l}$$

We find finally that

$$\Delta_{u,N}(u) f(x^R) = 4\vec{u}^2 \Delta_{x^R,n}(x^R) \tag{19}$$

We can also prove $\quad \Delta_{u,N}(u) f(x^L) = 4\vec{u}^2 \Delta_{x^L,n}(x^L) \tag{20}$

We conclude that the solutions of the Laplacian $\Delta_{u,N}(u)$ in the particular cases where N=2, 4, 8 and 16 are the generalization of the elements of the matrix representation of the group SU(2), or the bispherical harmonic function of this group. Moreover the research of the solutions of these Laplacians imposes a suitable parameterization of the variables $\{u\}$, what we will do in the following paragraph.

### 3.2. Parameterization of the transformation $R^8 \to R^5$

It is well known that the transformation of Cayley-Klein [3]

$$u_1 = \sqrt{r} \sin\frac{\theta}{2} \cos(\psi - \varphi), \; u_2 = \sqrt{r} \sin\frac{\theta}{2} \sin(\psi - \varphi),$$
$$u_3 = \sqrt{r} \cos\frac{\theta}{2} \cos(\psi + \varphi), \; u_4 = \sqrt{r} \cos\frac{\theta}{2} \sin(\psi + \varphi) \tag{21}$$

Is the parameterization of the application $R^4 \to R^3$ with $\psi$, $\theta$ and $\varphi$ are Euler's angles. This parameterization is due to the fact that the measure of integration on $S_3$ must also be valid on $S_2$.

The transformations of Hurwitz are quadratic transformations and noninvertible what causes the difficulty for the parameterization. Knowing that the last column in the matrix of rotation of SO(5) is the component of the unit vectors of $R^5$ we can overcome this difficulty, thus from the expression (11) of [11] we find that

With
$$\begin{pmatrix} x & y \\ -\bar{y} & \bar{x} \end{pmatrix} = 2\begin{pmatrix} \bar{v}_1 & -v_2 \\ \bar{v}_2 & v_1 \end{pmatrix}\begin{pmatrix} v_3 & v_4 \\ -\bar{v}_4 & \bar{v}_3 \end{pmatrix}$$
$$x = x_1^L + ix_2^L, \quad y = x_3^L + ix_4^L,$$
$$v_1 = u_1 + iu_2, \quad v_2 = u_3 + iu_4$$
$$v_3 = u_5 + iu_6, \quad v_4 = u_7 + iu_8$$
(22)

We obtain
$$x_5^L = (v_1\bar{v}_1 + v_2\bar{v}_2) - (v_3\bar{v}_3 + v_4\bar{v}_4)$$
$$x_1^L + ix_2^L = 2(\bar{v}_1 v_3 + v_2\bar{v}_4), \quad x_3^L + ix_4^L = 2(\bar{v}_1 v_4 - v_2\bar{v}_3),$$
$$x^L{}_5 = |v_1|^2 + |v_2|^2 - |v_3|^2 - |v_4|^2.$$

If we put
$$v_1 = \sqrt{r}\sin\frac{\chi}{2}w_1, \quad v_2 = \sqrt{r}\sin\frac{\chi}{2}w_2,$$
$$v_3 = \sqrt{r}\cos\frac{\chi}{2}w_3, \quad v_4 = \sqrt{r}\cos\frac{\chi}{2}w_4.$$

We deduce
$$x_1^L + ix_2^L = r\sin\chi(\bar{w}_1 w_3 + w_2\bar{w}_4), \quad x_3^L + ix_4^L = r\sin\chi(\bar{w}_1 w_4 - w_2\bar{w}_3),$$
$$x^L{}_5 = r\cos\chi$$
(23)

For the determination of $\{w_i\}$ we must find a transformation of which the number of parameters is 6. We will consider for that the representation matrix of SO (4) [9, 13]:

$$R_4 = \begin{pmatrix} t_1 & t_2 & 0 & 0 \\ -\bar{t}_2 & \bar{t}_1 & 0 & 0 \\ 0 & 0 & t_1 & t_2 \\ 0 & 0 & -\bar{t}_2 & \bar{t}_1 \end{pmatrix}\begin{pmatrix} t_5 & 0 & 0 & 0 \\ 0 & \bar{t}_5 & 0 & 0 \\ 0 & 0 & \bar{t}_5 & 0 \\ 0 & 0 & 0 & t_5 \end{pmatrix}\begin{pmatrix} t_3 & t_4 & 0 & 0 \\ -\bar{t}_4 & \bar{t}_3 & 0 & 0 \\ 0 & 0 & t_3 & t_4 \\ 0 & 0 & -\bar{t}_4 & \bar{t}_3 \end{pmatrix}$$

$$= \begin{pmatrix} w_1 & w_2 & 0 & 0 \\ -\bar{w}_2 & \bar{w}_1 & 0 & 0 \\ 0 & 0 & w_3 & w_4 \\ 0 & 0 & -\bar{w}_4 & \bar{w}_3 \end{pmatrix}$$

$$\begin{pmatrix} w_1 & w_2 \\ -\bar{w}_2 & \bar{w}_1 \end{pmatrix} = \begin{pmatrix} t_1 & t_2 \\ -\bar{t}_2 & \bar{t}_1 \end{pmatrix}\begin{pmatrix} t_5 & 0 \\ 0 & \bar{t}_5 \end{pmatrix}\begin{pmatrix} t_3 & t_4 \\ -\bar{t}_4 & \bar{t}_3 \end{pmatrix},$$

$$\begin{pmatrix} w_3 & w_4 \\ -\bar{w}_4 & \bar{w}_3 \end{pmatrix} = \begin{pmatrix} t_1 & t_2 \\ -\bar{t}_2 & \bar{t}_1 \end{pmatrix}\begin{pmatrix} \bar{t}_5 & 0 \\ 0 & t_5 \end{pmatrix}\begin{pmatrix} t_3 & t_4 \\ -\bar{t}_4 & \bar{t}_3 \end{pmatrix}$$
(24)

$$t_1 = e^{-\frac{i}{2}(\varphi+\psi)} \cos(\frac{\theta}{2}), \quad t_2 = e^{-\frac{i}{2}(\varphi-\psi)} \sin(\frac{\theta}{2})$$

$$t_3 = e^{-\frac{i}{2}(\varphi'+\psi')} \cos(\frac{\theta'}{2}), \quad t_4 = e^{-\frac{i}{2}(\varphi'-\psi')} \sin(\frac{\theta'}{2})$$

$$t_5 = e^{-\frac{\chi}{2}}, \quad \varphi = 0.$$

We have for the left side

$$\begin{pmatrix} \overline{w_1} & -w_2 \\ \overline{w_2} & w_1 \end{pmatrix} \begin{pmatrix} w_3 & w_4 \\ -\overline{w_4} & \overline{w_3} \end{pmatrix} = \begin{pmatrix} \overline{w_1}w_3 + w_2\overline{w_4} & \overline{w_1}w_4 - w_2\overline{w_3} \\ \overline{w_2}w_3 - w_1\overline{w_4} & \overline{w_2}w_4 + w_1\overline{w_3} \end{pmatrix} \quad (25)$$

And for the right side [5]

$$\begin{pmatrix} w_1 & w_2 \\ -\overline{w_2} & \overline{w_1} \end{pmatrix} \begin{pmatrix} \overline{w_3} & -w_4 \\ \overline{w_4} & w_3 \end{pmatrix} = \begin{pmatrix} w_1\overline{w_3} + w_2\overline{w_4} & -w_1 w_4 + w_2 \overline{w_3} \\ -\overline{w_2}\overline{w_3} + \overline{w_1}\overline{w_4} & \overline{w_2}w_4 + \overline{w_1}w_3 \end{pmatrix}$$

Introduce (23) in the expression (24) we find

$$\overline{w_1}w_3 + w_2\overline{w_4} = \cos\chi + i\sin\chi\cos\theta'$$
$$\overline{w_1}w_4 - w_2\overline{w_3} = i\sin\chi\sin\theta'\cos\psi' + \sin\chi\sin\theta'\sin\psi' \quad (26)$$

Finally we find the expression of last row:

$$x_1^L = r\sin\eta\cos\chi, \qquad x_2^L = r\sin\eta\sin\chi\cos\theta'$$
$$x_3^L = r\sin\eta\sin\chi\sin\theta'\sin\psi', \qquad x_4^L = r\sin\eta\sin\chi\sin\theta'\cos\psi'$$
$$x_5^L = r\cos\eta$$

Therefore we deduce from (23) and (24) the parameterization

$$z_1 = u_1 + iu_2 = \sqrt{r}\cos(\frac{\eta}{2})(t_1 t_3 e^{i\frac{\chi}{2}} - t_2 \bar{t}_4 e^{-i\frac{\chi}{2}}), \quad z_2 = u_3 + iu_4 = \sqrt{r}\cos(\frac{\eta}{2})(t_1 t_4 e^{i\frac{\chi}{2}} + t_2 \bar{t}_3 e^{-i\frac{\chi}{2}}),$$

$$z_3 = u_5 + iu_6 = \sqrt{r}\sin(\frac{\eta}{2})(t_1 t_3 e^{-i\frac{\chi}{2}} - t_2 \bar{t}_4 e^{i\frac{\chi}{2}}), \quad z_4 = u_7 + iu_8 = \sqrt{r}\sin(\frac{\eta}{2})(t_1 t_4 e^{-i\frac{\chi}{2}} + t_2 \bar{t}_3 e^{i\frac{\chi}{2}}),$$

### 3.3 Existence of the bispherical harmonic functions

For the clearness of the presentation we consider first the case of rotation and we adopt the general notations used in group theory [10, 13] for the classical groups what makes generalization apparent. It's well known from the theory of angular momentum [14] that there are two spherical harmonics the left and the right:

$$\left\langle \begin{matrix}[l]\\(0)\end{matrix} \middle| R(\psi\theta\varphi) \middle| \begin{matrix}[l]\\(m)\end{matrix} \right\rangle = (-1)^m \left(\frac{4\pi}{2l+1}\right)^{\frac{1}{2}} Y_{lm}(\theta\varphi) \text{ and } \left\langle \begin{matrix}[l]\\(m)\end{matrix} \middle| R(\psi\theta\varphi) \middle| \begin{matrix}[l]\\(0)\end{matrix} \right\rangle = \left(\frac{4\pi}{2l+1}\right) Y_{lm}(\psi\theta)$$

The product may be written as

$$\left\langle \begin{matrix}[l_1]\\(0)\end{matrix}\middle| R_1(\psi\theta\varphi)\middle| \begin{matrix}[l_1]\\(m_1)\end{matrix}\right\rangle \left\langle \begin{matrix}[l_2]\\(m_2)\end{matrix}\middle| R_2(\psi\theta\varphi)\middle| \begin{matrix}[l_2]\\(0)\end{matrix}\right\rangle$$
$$= \left\langle \begin{matrix}[l_1 l_2]\\(0 m_2)\end{matrix}\middle| R_{12}(\psi\theta\varphi)\middle| \begin{matrix}[l_1 l_2]\\(m_1 0)\end{matrix}\right\rangle \tag{16}$$

With $R_{12}$ is the product rotation $R_1 R_2$.

The coupling of angular momentum of $\left|\begin{matrix}[l_1 l_2]\\(m_1 0)\end{matrix}\right\rangle$ is given by

$$\left|\begin{matrix}[l_1 l_2]\\(m_1 0)\end{matrix}\right\rangle = \sum_l \left\{\left\langle \begin{matrix}[(l_1 l_2)l]\\(m_1)\end{matrix}\middle\| \begin{matrix}[l_1 l_2]\\(m_1 0)\end{matrix}\right\rangle\right\} \left|\begin{matrix}[(l_1 l_2)l]\\(m_1)\end{matrix}\right\rangle$$

With $\left\langle \begin{matrix}[(l_1 l_2)l]\\(m_2 0)\end{matrix}\middle\| \begin{matrix}[(l_1 l_2)l]\\(m_2)\end{matrix}\right\rangle$ is the Clebsh-Gordan coefficient.

Using (16) we deduce that

$$\left\langle \begin{matrix}[l_1 l_2]\\(0 m_2)\end{matrix}\middle| R_{12}(\psi\theta\varphi)\middle| \begin{matrix}[l_1 l_2]\\(m_1 0)\end{matrix}\right\rangle = \sum_l \left\{\left\langle \begin{matrix}[(l_1 l_2)l]\\(m_2 0)\end{matrix}\middle\| \begin{matrix}[(l_1 l_2)l]\\(m_2)\end{matrix}\right\rangle\right\} \times \left\{\left\langle \begin{matrix}[(l_1 l_2)l]\\(0 m_1)\end{matrix}\middle\| \begin{matrix}[(l_1 l_2)l]\\(m_2)\end{matrix}\right\rangle\right\}$$
$$\times \left\langle \begin{matrix}[(l_1 l_2)l]\\(m_2)\end{matrix}\middle| R_{12}(\psi\theta\varphi)\middle| \begin{matrix}[(l_1 l_2)l]\\(m_1)\end{matrix}\right\rangle . \tag{17}$$

The element of the matrix of rotation

$$\left\langle \begin{matrix}[(l_1 l_2)l]\\(m_2)\end{matrix}\middle| R_{12}(\psi\theta\varphi)\middle| \begin{matrix}[(l_1 l_2)l]\\(m_1)\end{matrix}\right\rangle = D^l_{(m_2, m_1)}(\psi\theta\varphi), l = 0,1,..., \tag{18}$$

is the bispherical harmonic function of SO(3) which can be easily generalized to the classical groups.

### 3.4 The coordinates of the bispherical harmonic functions

We notice that the spherical coordinates on the sphere 'left' $S^L_{n-1}$ is $(\theta_1,...,\theta_{n-2},\theta_{n-1})$.

The spherical coordinates on the sphere 'right' is $S^R_{n-1}$ $(\theta'_1,...,\theta'_{n-2},\theta_{n-1})$ and on the sphere $S_{2(n-1)}$ is $(\theta_1,...,\theta_{n-2},\theta_{n-1},\theta'_{n-2},...,\theta'_1)$.

The points:

$$x = (x_1,...,x_{n-2},x_n) \in S^L_{n-1} \text{ and } x' = (x'_1,...,x'_{n-2},x'_n) \in S^R_{n-1},$$

can be represented as

$$x_1 = \sin\theta_{n-1}\sin\theta_{n-2}...\sin\theta_1, \quad x'_1 = \sin\theta_{n-1}\sin\theta'_{n-2}...\sin\theta'_1$$
$$x_2 = \sin\theta_{n-1}\sin\theta_{n-2}...\cos\theta_1, \quad x'_2 = \sin\theta_{n-1}\sin\theta'_{n-2}...\cos\theta'_1$$
$$............................, \quad ............................$$
$$x_{n-2} = \sin\theta_{n-1}\cos\theta_{n-2}, \quad x'_{n-2} = \sin\theta_{n-1}\cos\theta'_{n-2}$$
$$x_{n-1} = \cos\theta_{n-1}, \quad x'_{n-1} = \cos\theta_{n-1}$$

Owing to the fact that the spherical functions and the bispherical functions are elements of the matrix of rotation of SO(N), it results from that the invariant measure on the sphere $S_{2(n-1)}$ must be the integration measure on $S^L_{n-1}$ and $S^R_{n-1}$. This result and the generalization of (21-23) allows us to deduce the coordinates of a point on the sphere $S_{2(n-1)}$

$$x_1 = \sin\frac{\theta_{n-1}}{2}\sin\theta_{n-2}\ldots\sin\theta_1, \quad x_{n+1} = \cos\frac{\theta_{n-1}}{2}\cos\theta'_{n-2}$$

$$x_2 = \sin\frac{\theta_{n-1}}{2}\sin\theta_{n-2}\ldots\cos\theta_1, \quad \ldots\ldots\ldots\ldots\ldots\ldots\ldots$$

$$\ldots\ldots\ldots\ldots\ldots\ldots\ldots, \quad x_{2n-1} = \cos\frac{\theta_{n-1}}{2}\sin\theta'_{n-2}\ldots\cos\theta_1$$

$$x_{n-1} = \sin\frac{\theta_{n-1}}{2}\cos\theta_{n-2}, \quad x_{2(n-1)} = \cos\frac{\theta_{n-1}}{2}\sin\theta'_{n-2}\ldots\sin\theta_1$$

## 4. The generating matrices and the Cartan-Weyl basis

We start with the link of the Cartan-Weyl basis for the group SO(3),SO(4) and SO(5) with Hurwitz matrices. We use the traditional notations of adjoint representations of these groups (Appendix). In the general case it is easier to use the notations of the adjoint representations [10, 16] of the group SO(n) in term of the matrix

$$\Sigma_{ij} = (\delta_{ik}\delta_{jl} - \delta_{jk}\delta_{il}), \quad i,j,k,l = 1,\ldots,n$$

### 4.1. The Hurwitz's matrix and the generating matrices
#### 4.1.1 Generating matrices of SO(3)
For n= 2 we have

$$H_2 = u_1 I + u_2 \widehat{S}_3$$

For n= 3 we must add to the case n =2 $\widehat{S}_1$ and $\widehat{S}_2$.

$$H_3 = u_1 I + u_2 \widehat{S}_3 + u_3 \widehat{S}_2 + u_4 \widehat{S}_1$$

#### 4.1.2 Generating matrices of SO(4)
For n= 4 we obtain by Cayley transformation two orthogonal matrices, the left and the right:

$$H_4^1 = u_1 I + u_2 \widehat{S}_3 + u_3 \widehat{S}_2 + u_4 \widehat{S}_1$$
$$H_4^2 = u_1 I + u_2 \widehat{T}_3 + u_3 \widehat{T}_2 + u_4 T_1 \tag{27}$$

S and T are two commuting spins which generate SO(4) transformations.

#### 4.1.3 Generating matrices of SO(5)
For n= 5 we must add only $\widehat{U}_1, \widehat{U}_2, \widehat{V}_1, \widehat{V}_2$ to the above-mentioned matrices and we write:

$$H_5^1 = u_1 I + u_2 \hat{S}_3 + u_3 \hat{S}_2 + u_4 \hat{S}_1 + u_5 \hat{U}_1 + u_6 \hat{U}_2 + u_7 \hat{V}_1 + u_8 \hat{V}_2$$

$$= \begin{pmatrix} u_1 & u_2 & u_3 & u_4 & u_5 \\ -u_2 & u_1 & u_4 & -u_3 & u_6 \\ -u_3 & -u_4 & u_1 & u_2 & u_7 \\ -u_4 & u_3 & -u_2 & u_1 & u_8 \\ -u_5 & -u_6 & -u_7 & -u_8 & u_1 \end{pmatrix} \quad (28)$$

We must change S by T to obtain the other matrix $H_5^2$.

### 4.2. Generating matrices of SO($2^n$)

We treat first n= 8 and then the general case.

### 4.2.1 Generating matrices of SO(8)

For n= 8 we obtain by Cayley transformation two matrix from $H_5^1$ and two matrix from $H_5^2$ and after calculations we find that three of these matrices are orthogonal. Moreover these matrices do not generate the Cartan basis what we will do in what follows.

The number of generating matrices is four thus we must group the elements of the adjoint representation in four groups, work already carry out by many authors [15-18]. We start from the orthogonal Hurwitz matrix $H_8^1$ writing in term of adjoint representation $\{\Sigma_{ij}\}$:

$$H_8^1 = \begin{pmatrix} u_1 & -u_2 & -u_3 & -u_4 & u_5 & -u_6 & -u_7 & -u_8 \\ u_2 & u_1 & u_4 & -u_3 & u_6 & u_5 & -u_8 & u_7 \\ u_3 & -u_4 & u_1 & u_2 & u_7 & u_8 & u_5 & -u_6 \\ u_4 & u_3 & -u_2 & u_1 & u_8 & -u_7 & u_6 & u_5 \\ -u_5 & -u_6 & -u_7 & -u_8 & u_1 & u_2 & u_3 & u_4 \\ u_6 & -u_5 & -u_8 & u_7 & -u_2 & u_1 & -u_4 & u_3 \\ u_7 & u_8 & -u_5 & -u_6 & -u_3 & u_4 & u_1 & -u_2 \\ u_8 & -u_7 & u_6 & -u_5 & -u_4 & -u_3 & u_2 & u_1 \end{pmatrix} \quad (29)$$

$$\begin{aligned}
&= u_1[+\Sigma_{11} + \Sigma_{22} + \Sigma_{33} + \Sigma_{44}] + u_5[+\Sigma_{15} + \Sigma_{26} + \Sigma_{37} + \Sigma_{48}] + \\
&\quad u_2[-\Sigma_{12} + \Sigma_{34} + \Sigma_{56} - \Sigma_{78}] + u_6[-\Sigma_{16} + \Sigma_{25} - \Sigma_{38} + \Sigma_{47}] + \\
&\quad u_3[-\Sigma_{13} - \Sigma_{24} + \Sigma_{57} + \Sigma_{68}] + u_7[-\Sigma_{17} + \Sigma_{28} + \Sigma_{35} - \Sigma_{46}] + \\
&\quad u_4[-\Sigma_{14} + \Sigma_{23} - \Sigma_{67} + \Sigma_{58}] + u_8[-\Sigma_{18} - \Sigma_{27} + \Sigma_{36} + \Sigma_{45}]
\end{aligned} \quad (30)$$

We remark that the matrices (4, 4)

$$\begin{pmatrix} 1 & 1 & 1 & 1 \\ -1 & 1 & 1 & -1 \\ -1 & -1 & 1 & 1 \\ -1 & 1 & -1 & 1 \end{pmatrix}, \quad \begin{pmatrix} 1 & 1 & 1 & 1 \\ -1 & 1 & -1 & 1 \\ -1 & 1 & 1 & -1 \\ -1 & -1 & 1 & 1 \end{pmatrix}$$

formed from the coefficients of Σ are orthogonal therefore we find the matrices $H_8^i, i = 2 \cdots 8$ by cyclic permutation of the row of these coefficients in (30).

**4.2.1 Generating matrices for the general case**

We determine the anti-symmetric matrices $(H_{2^n}), n = 4, 5, \ldots,$ by recurrence using the method exposed in [11] part 6.3. We write this matrix in the form (29) like above. Knowing that the matrices of coefficients are not obligatorily different from zero and owing to the fact that the generating matrix is not unique we can solve this problem by changing the coefficients by the coefficients of the well known Hadamard's matrix which is defined by:

$$H_1^S = \begin{pmatrix} 1 & 1 \\ 1 & -1 \end{pmatrix}$$

And

$$H_n^S = H_1^S \otimes H_{n-1}^S = \begin{pmatrix} H_{n-1}^S & H_{n-1}^S \\ H_{n-1}^S & -H_{n-1}^S \end{pmatrix}$$

By cyclic permutation of the row of the matrix of coefficients we obtain the generating matrices.

**5. Appendix: The generators of SO (5) groups**

The ten generators of SO (5), the group of rotations in five dimensions [19-20], may be taken as $L_{ij} = -L_{ji}; i \neq j; i, j = 1, \cdots, 5$. $L_{ij}$ what generates rotations in the ij plane.

The commutations rules for the L's are

$$L_{ij} = -i(x_i \frac{\partial}{\partial x_j} - x_j \frac{\partial}{\partial x_i})$$

$$[L_{ij}, L_{kl}] = i(\delta_{ik} L_{jl} + \delta_{jk} L_{ik} - \delta_{il} L_{jk} - \delta_{jk} L_{il})$$

We define

$$[E_{ij}, E_{kl}] = \delta_{jl} E_{il} - \delta_{il} E_{kj}$$

$$E_{ij} = \sum_{p=1}^{2} a_i^{+p} a_j^p, \quad (i, j = 1, \ldots, 4)$$

$a_i^+$ and $a_i$ are the operators of creation and destruction of the harmonic oscillator.

$$S_1 = L_{23} + L_{14}, \quad S_2 = L_{31} + L_{24}, \quad S_3 = L_{12} + L_{34}$$
$$T_1 = L_{23} - L_{14}, \quad T_2 = L_{31} - L_{24}, \quad T_3 = L_{12} - L_{34}$$
$$U_1 = L_{15}, \quad U_2 = L_{25}, \quad U_\pm = L_{15} \pm iL_{25}$$
$$V_1 = L_{35}, \quad V_2 = L_{45}, \quad V_\pm = L_{25} \pm iL_{45}$$
$$S_\pm = S_1 \pm S_2, \quad T_\pm = T_1 \pm T_2$$

Then S and T are two commuting spins which generate SO(4) transformations. We put a hat to the adjoint representations.